# Thermodynamic derivation of reciprocal relations

Elias P. Gyftopoulos
Massachusetts Institute of Technology
77 Massachusetts Avenue - Room 24-111
Cambridge, Massachusetts 02139 USA
epgyft@aol.com

Reciprocal relations correlate fairly accurately a great variety of experimental results. Nevertheless, the concepts of statistical fluctuations, and microscopic reversibility – the bases of the accepted proof of the relations by Onsager – are illusory and faulty, and contradict the foundations of the science of thermodynamics. The definitions, postulates, and main theorems of thermodynamics are briefly presented. It is shown beyond a shadow of a doubt that thermodynamics is a nonstatistical science that applies to all systems (both macroscopic, and microscopic, including systems that consist either of only one structureless particle, or only one spin), to all states (both thermodynamic or stable equilibrium, and not stable equilbrium), and that includes entropy as a well defined, intrinsic, nonstatistical property of any system in any state, at any instant in time. In the light of this novel conception of thermodynamics, we find that reciprocal relations result from a well known mathematical theorem, to wit, given a well behaved analytic function $f(a_1, a_2, ..., a_m)$ then $\partial^2 f / \partial a_i \partial a_j \equiv \partial^2 f / \partial a_j \partial a_i$ for i, j = 1, 2, …, m.

## INTRODUCTION

On the occasion of the commemoration of the 100th birthday of Nobel Prize winner Lars Onsager, we honor him for his pioneering attempts to justify "reciprocal relations in irreversible processes" [1, 2] by using statistical classical mechanics, and what he called "the concept of microscopic reversibility", a concept that includes the consideration of past, present, and future times, and averages of spontaneous statistical fluctuations beginning and ending in a thermodynamic or stable equilibrium state. His work has stimulated many related studies by Meixner [3-6], Nobel Prize winner Ilya Prigogine [7], Nobel Prize winner Peter Dennis Mitchell [8], de Groot and Mazur [9], Haase [10], Katchalsky and Curran [11], and many other scientists. In these studies, reciprocal relations are fairly accurately verified by experimental results obtained from thermoelectric devices, electrolytic cells, and other processes. In fact, some of these relations were empirically recognized even before Onsager provided his theoretical explanation. This experimental verification is viewed by practically all scientists and engineers as one incontrovertible proof of the validity of statistical classical mechanics, and microscopic reversibility.

Despite the satisfactory experimental verification of reciprocal relations, it behooves us to reexamine their theoretical justification in the light of our current understandings of the concept of time, and of the science of thermodynamics.

Specifically, Einstein – the Man of the 20th century – says [12]: "For us loyal physicists, this separation between past, present, and future holds only the value of an illusion, tenacious as it may be". And yet, this separation plays a key role in the evaluation of cross-correlations of the statistical fluctuations presumed by Onsager, fluctuations which temporarily reduce the entropy without compensation and, therefore, violate the laws of thermodynamics. Said differently, fluctuations violate the definition of an equilibrium state that is globally stable, a definition which is more general than that used by Poincaré and Liapunov in mechanics for local stability.

PACS number(s): 05.70.Ln, 03.65.-w, 05.30.-d

---

Again, we have the perennial question, best posited by Feynman. In trying to understand entropy and irreversibility, he considers the mixing of white and black molecules, and comments [13]: "Gradually the whites worm their way, by accident, across into the space of the blacks, and the blacks worm their way, by accident, into the space of the whites. Clearly, this is an irreversible process in the real world, and ought to involve an increase in the entropy." The conclusion about the entropy of mixing is correct and real. However, the cause of this reality is not the mixing of the kinds of molecules considered by Feynman because he tacitly assumes that each molecule has a fixed spatial shape, and if this assumption is valid then there is no entropy of mixing [14], the process is reversible. The cause of irreversibility – the entropy of mixing – is the change in volume in which each colored molecule is confined, and the ensuing change of the quantum mechanical spatial shape of each molecule [15, 16].

Again, a different but pervasive answer which is shared by practically all colleagues that profess and nurture the science of statistical classical mechanics is given by Maxwell. He says [17]: "In dealing with masses of matter while we do not perceive the individual molecules, we are compelled to adopt what I have described as the statistical method of calculation, and to abandon the strict dynamical method, in which we follow every motion by the calculus". But is it possible to have irreversible processes just because we do not know how to solve numerically difficult problems? And conversely, would the concept of irreversibility disappear from our textbooks, our analyses, and the myriad of irreversible processes that we encounter in practice, if our analytical and computational skills improve so much that we can follow every motion by the calculus? Our answer to each of these rhetorical questions is an unqualified no. Our answer is so emphatic because in the case of mixing perfect gases, the spontaneous entropy increase is based on the behavior of each individual molecule and, therefore, can be followed precisely by the calculus alluded to by Maxwell [14].

To be sure, the statistical method of calculation yields many numerically correct answers, in the same sense that the geocentric model of our solar system yields many numerically correct answers. However, the effects of irreversibility are real, and do not depend on our ability to solve difficult mathematical problems, in the same sense that the stationary electronic energy eigenstates of a heavy element in the periodic table are real despite our difficulties to calculate the precise energy eigenvalues, and energy eigenfunctions.

Even though tremendous progress has been made over the past two and a half centuries in improving the efficiency of various processes, that is, in reducing the spontaneous generation of entropy, irreversibility continues to result in irreparable exhaustion of energy resources, and to cause the degradation of the earth's environment in ways that are energetically costly and hard to curtail, and perhaps impossible to sustain.

In response to dilemmas such as discussed in the preceding remarks, a small group of faculty and students at MIT has conceived of and elaborated on two intimately interrrelated and trailblazing theoretical interpretations of physical phenomena.

One consists in recognizing that what Feynman calls [13] "the laws of physics" are correct but incomplete. Specifically, the laws of quantum mechanics that involve only wave functions or projectors are incomplete because they apply only to processes that are unitary in time, and therefore are reversible. But not all reversible phenomena are unitary, and not all phenomena are reversible. We have addressed this issue by developing a nonstatistical unified quantum theory of mechanics and thermodynamics [18] that completes the "laws of physics", and that includes a nonlinear equation of motion [19, 20]. In the unified theory, Feynman's laws of physics are a special case, zero entropy physics, and the Onsager reciprocity and dispersion - dissipation relations are derived without resort to statistical physics, and microscopic reversibility [21].

In sharp contrast to what is presented in practically every textbook on physics and/or thermodynamics, our second regularization of physical phenomena consists of a novel, non-quantal exposition of thermodynamics as a nonstatistical science that applies to all systems (both macroscopic, and microscopic, including systems that consist of either only one structureless particle, or only one spin), and to all states (both thermodynamic or stable equilibrium, and not stable equilibrium) [22]. An important proven theorem (not law) of this nonstatistical exposition is the existence of entropy as a well defined (without ambiguities, circular arguments, and the concepts of heat and temperature) intrinsic, nonstatistical property of any system, in any state, at any instant in time. A brief summary of this novel exposition is given in the Appendix.

On the basis of the novel exposition, which does not involve statistics, fluctuations, and microscopic reversibility, in what follows we derive the reciprocal relations by using the same flow processes, the same linear approximations, and the same experimental results as all previous investigators, and the fact that a well behaved analytic function $f(a_1, a_2, ..., a_r)$ satisfies the identity $\partial^2 f / \partial a_i \partial a_j \equiv \partial^2 f / \partial a_j \partial a_i$ for i, j = 1, 2, ..., r.

## RECIPROCAL RELATIONS FOR SYSTEMS WITHOUT CHEMICAL REACTIONS

### General remarks

From the practical standpoint, the most important and interesting applications of thermodynamics involve systems passing through nonequilibrium states. For example, an energy conversion system, a chemical reactor, and a living entity and its life support processes are each a system in a nonequilibrium state evolving in time. In general, a rigorous analysis of a system passing through nonequilibrium states is cumbersome because it requires both a very large number of independent variables – a very large number of independent properties at each instant in time – and the complete equation of motion alluded to earlier.

Inspired by the success of numerical correlations of experimental data by means of Fourier's law, Fick's law, and Ohm's law, we approximate nonequilibrium states by the following model.

(i) We consider a one-dimensional system $A$ that experiences no chemical reactions, has a fixed volume, and is in a steady state (a state that does not vary in time) as a result of interactions with two reservoirs I and II as shown in Figure 1, where $\tau$ is the inverse absolute temperature, and $\mu_i$ the chemical or electrochemical potential of the ith neutral or electrically charged constituent, respectively, for i = 1, 2, ..., r.

| Reservoir I | | Reservoir II |
|---|---|---|
| $\tau^I$ | System $A$ | $\tau^{II}$ |
| $\mu_1^I \tau^I$ | in a | $\mu_1^{II} \tau^{II}$ |
| $\mu_2^I \tau^I$ | steady state | $\mu_2^{II} \tau^{II}$ |
| . | | . |
| . | | . |
| . | | . |
| $\mu_r^I \tau^I$ | | $\mu_r^{II} \tau^{II}$ |

*Figure 1*. Schematic representation of system $A$ maintained in a steady state by two reservoirs.

(ii) We define a reservoir as a system that, to a very high degree of accuracy, satisfies the following specifications: (a) It passes through stable equilibrium states only; (b) In the course of finite changes of state at constant or varying values of its energy, amounts of constituents, and parameters (such as volume and/or an applied electric field), it departs infinitesimally from the condition of mutual stable equilibrium with a duplicate of itself that experiences no such changes; (c) At constant values of the amounts of constituents and parameters of each of two reservoirs initially in mutual stable equilibrium, energy can be transferred reversibly from one reservoir to the other at the expense of infinitesimal and negligible effects on any other system.

(iii) As a result of (i) and (ii), all time dependent changes occur only in reservoirs I and II. But because each reservoir passes only through stable equilibrium states, all properties and all time dependent changes can be interrelated by using the fundamental relation, namely, the function $S(E, n, V, \mathcal{E})$ which relates the entropy $S$ of stable equilibrium states of either reservoir I or II to energy $E$, amounts of r constituents $n_1, n_2, \ldots, n_r$ denoted by the vector $\mathbf{n}$, and the two parameters, volume $V$ and electrostatic field $\mathcal{E}$. The function $S(E, n, V, \mathcal{E})$ is shown to be [22] a well behaved analytic function of each of its independent variables $E, n, V, \mathcal{E}$, and therefore it is differentiable to all orders with respect to each of these variables. In particular, it is the function used for the rigorous definitions of temperature $T = 1/\tau$, chemical or electrochemical potential $\mu_i$, and pressure $p$ by the relations [22]

$$\tau = 1/T = (\partial S/\partial E)_{n,\beta}$$

$$-\mu_i \tau = (\partial S/\partial n_i)_{E,n,\beta} \quad (1)$$

$$p\tau = (\partial S/\partial V)_{E,n,\beta}$$

(iv) Because system $A$ is in a steady state, we can partition it into contiguous laminas of differential thickness dx, and assert that each lamina is in a steady state as shown in Figure 2. It is noteworthy, that partitioning is possible if and only if the system is *simple*. The definition, and characteristics of simple systems are discussed in [22].

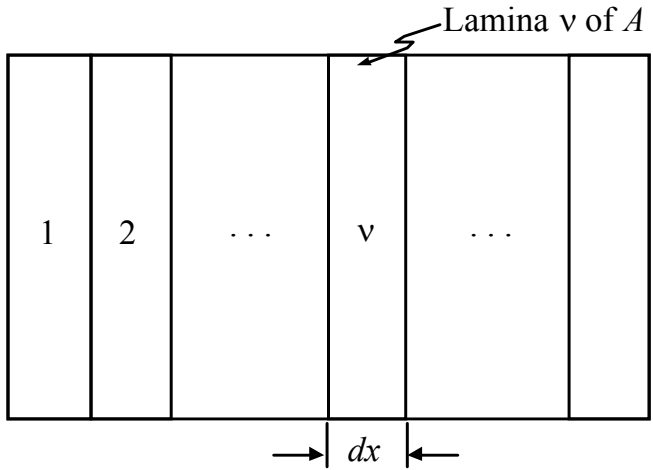

*Figure 2.* Schematic representation of laminated system $A$.

(v) Capitalizing on the excellent results that are obtained from the use of Fourier's, Fick's, and Ohm's laws, we can analyze each lamina as if it were in a steady state between two local reservoirs, as shown in Figure 3.

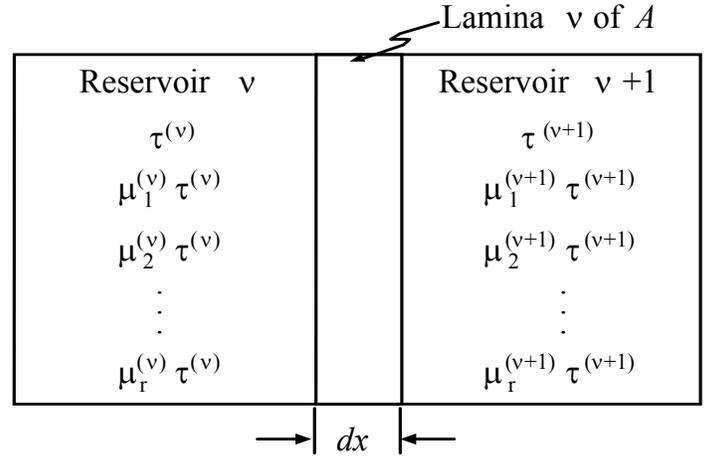

*Figure 3.* Schematic representation of the $\nu$-th lamina maintained in a steady state by reservoirs $\nu$ and $\nu+1$.

**Rate of entropy generation**

The rates of change of entropy of reservoirs $\nu$ and $\nu + 1$ per unit surface area of the lamina between interfaces $\nu$ and $\nu + 1$ are given by the relations

$$\dot{S}^{(\nu)} = \tau^{(\nu)} \dot{E}^{(\nu)} + \sum_{i=1}^{r} (-\mu_i^{(\nu)}) \dot{n}_i^{(\nu)} \quad (2)$$

$$\dot{S}^{(\nu+1)} = \tau^{(\nu+1)} \dot{E}^{(\nu+1)} + \sum_{i=1}^{r} (-\mu_i^{(\nu+1)}) \dot{n}_i^{(\nu+1)} \quad (3)$$

where superscripts $\nu$ and $\nu + 1$ refer to the interfaces $\nu$ and $\nu + 1$, respectively, and a dot indicates derivative with respect to time.

Because the lamina is in a steady state,

$$-\dot{E}^{(\nu)} = \dot{E}^{(\nu+1)} = J_\varepsilon \quad (4)$$

$$-\dot{n}_i^{(\nu)} = \dot{n}_i^{(\nu+1)} = J_i \quad \text{for} \quad i = 1, 2, \ldots, r \quad (5)$$

and, therefore, the rate of entropy generation per unit area of the face of the lamina is given by the relation

$$J_s = \dot{S}^{(\nu)} + \dot{S}^{(\nu+1)} = J_\varepsilon d\tau + \sum_{i=1}^{r} J_i d(-\mu_i \tau) \quad (6)$$

where

$$d\tau = \tau^{(\nu+1)} - \tau^{(\nu)} \quad (7)$$

$$d(-\mu_i \tau) = \mu_i^{(\nu)} \tau^{(\nu)} - \mu_i^{(\nu+1)} \tau^{(\nu+1)} \quad (8)$$

Moreover, the rate of entropy generation per unit lamina thickness is given by the relation

$$J_s' = J_\varepsilon \tau' + \sum_{i=1}^{r} J_i (-\mu_i \tau)' \quad (9)$$

where a prime indicates a derivative with respect to $x$. It is customary to interpret Eq. 6 or 9 as representing the rate of entropy generation in terms of the fluxes $J_\varepsilon, J_1, J_2, ..., J_r$, and the forces either $d\tau$ and $d(-\mu_i \tau)$, or $\tau'$ and $(-\mu_i \tau)'$ for i = 1, 2,..., r.

The preceding analysis can be readily extended to more than one dimension for both isotropic, and nonisotropic systems in steady states.

**Reciprocal relations**

For the elementary process shown in Figure 3, any steady state flux density $J_m$ can be expressed as a function of the form

$$J_m = J_m[\tau, (-\mu_1 \tau), (-\mu_2 \tau),..., \tau', (-\mu_1 \tau)', (-\mu_2 \tau)',...] \quad (10)$$

where m = $\varepsilon$, 1, 2,..., r, and for the sake of simplicity we omit the superscript $\nu$. In the form of Eq. 10, $J_m$ is said to depend on the variables that characterize reservoir $\nu$, and the forces $\tau', (-\mu_m \tau)'$ for m = $\varepsilon$, 1, 2,..., r evaluated at the interface $\nu$.

We can expand each $J_m$ into a Maclaurin series of the forces of the form

$$J_m = [\partial J_m / \partial \tau']_0 \tau'$$
$$+ \sum_{i=1}^{r} [\partial J_m / \partial (-\mu_i \tau)']_0 (-\mu_i \tau)' + [h.o.t.] \quad (11)$$

where subscript "0" denotes that in the evaluation of each partial derivative, the value of each force $d\tau$ and $d(-\mu_i \tau)$ or $\tau'$ and $d(-\mu_i \tau)'$ for all i is taken equal to zero, and [h.o.t.] stands for higher order terms. The term
$J_m[\tau, (-\mu_1 \tau),..., \tau' = 0, (-\mu_1 \tau)' = 0,...]$
does not appear in the Maclaurin series because if the forces are equal to zero, then the fluxes are equal to zero. Moreover, a large number of experiments indicate that each flux can be fairly accurately represented by an expression of the form of Eq. 11 without the higher order terms.

Without any loss of generality, in what follows we consider a steady state process that involves only three fluxes, $J_\varepsilon$ of energy, $J_1$ of neutral particles, and $J_2$ of electrically charged particles, and write the equations for the rate of entropy generation, and the fluxes in the forms

$$J_s' = J_\varepsilon \tau' + J_1 (-\mu_1 \tau)' + J_2 (-\mu_2 \tau)' \quad (12)$$

$$J_\varepsilon = L_{\varepsilon\varepsilon} \tau' + L_{\varepsilon 1} (-\mu_1 \tau)' + L_{\varepsilon 2} (-\mu_2 \tau)' \quad (13)$$

$$J_1 = L_{1\varepsilon} \tau' + L_{11} (-\mu_1 \tau)' + L_{12} (-\mu_2 \tau)' \quad (14)$$

$$J_2 = L_{2\varepsilon} \tau' + L_{21} (-\mu_1 \tau)' + L_{22} (-\mu_2 \tau)' \quad (15)$$

Using equation 12 we find

$$J_\varepsilon = \partial J_s' / \partial \tau'$$
$$J_1 = \partial J_s' / \partial (-\mu_1 \tau)' \quad (16)$$
$$J_2 = \partial J_s' / \partial (-\mu_2 \tau)'$$

Using the fact that for a well behaved analytic function $f(a_1, a_2,..., a_r)$

$$\partial^2 f / \partial a_i da_j \equiv \partial^2 f / da_j da_i \quad \text{for} \quad i, j = 1, 2, ..., r \quad (17)$$

and Eqs. 12 to 15, we find

$$L_{\varepsilon 1} = \frac{\partial J_\varepsilon}{\partial(-\mu_1\tau)'} = \frac{\partial^2 J'_s}{\partial(-\mu_1\tau)'\partial\tau'}; \tag{18a}$$

$$L_{1\varepsilon} = \frac{\partial J_1}{\partial \tau'} = \frac{\partial^2 J'_s}{\partial \tau'\partial(-\mu_1\tau)'} = L_{\varepsilon 1} \tag{18b}$$

$$L_{\varepsilon 2} = \frac{\partial J'_\varepsilon}{\partial(-\mu_2\tau)'} = \frac{\partial^2 J'_s}{\partial(-\mu_2\tau)'\partial\tau'} \tag{19a}$$

$$L_{2\varepsilon} = \frac{\partial J_2}{\partial \tau'} = \frac{\partial^2 J'_s}{\partial \tau'\partial(-\mu_2\tau)'} = L_{\varepsilon 2} \tag{19b}$$

$$L_{12} = \frac{\partial J_1}{\partial(-\mu_2\tau)'} = \frac{\partial^2 J'_s}{\partial(-\mu_2\tau)'\partial(-\mu_1\tau)'} \tag{20a}$$

$$L_{21} = \frac{\partial J_2}{\partial(-\mu\tau_1)'} = \frac{\partial^2 J'_s}{\partial(-\mu_1\tau)'\partial(-\mu_2\tau)'} = L_{12} \tag{20b}$$

that is, we prove the reciprocal relations without resorting to statistical fluctuations, the concepts of past, present, and future, and the idea of microscopic reversibility which are not valid for nonzero entropy physics of any system (both macroscopic, and microscopic), in any nonzero entropy state (both stable equilibrium, and not stable equilibrium).

## CONCLUSION

For the first time in the long history of the problem, reciprocal relations are not derived as consequences of the statistical classical mechanics interpretation of thermodynamics, fluctuations, and microscopic reversibility. They are proven on the basis of a rigorous, unambiguous, and noncircular exposition of thermodynamics as a nonstatistical theory of physics that applies to all systems, and all states, and that includes entropy as an intrinsic property of the constituents of a system, in the same sense that rest mass is an intrinsic property of the constituents of a system.

In this presentation, we discuss reciprocal relations of a system that experiences no chemical reactions. Thermodynamic derivations of conditions for chemical equilibrium and of Onsager reciprocal relations for chemical reactors are discussed in a companion paper [23].

## APPENDIX: A NEW EXPOSITION OF THERMODYNAMICS

### A.1 Basic concepts

#### A.1.1 General remarks

Many scientists and engineers have expressed concerns about the completeness and clarity of the usual expositions of thermodynamics. For example, in the preface of his book *Concepts of Thermodynamics*, Obert writes [24] "Most teachers will agree that the subject of engineering thermodynamics is confusing to the student despite the simplicity of the usual undergraduate presentation." Again, Tisza states [25] "The motivation for choosing a point of departure for a derivation is evidently subject to more ambiguity than the technicalities of the derivation….In contrast to errors in experimental and mathematical techniques, awkward and incorrect points of departure have a chance to survive for a long time."

In response to numerous such concerns, Gyftopoulos and Beretta [22] have composed an exposition in which all basic concepts of thermodynamics are defined completely, without ambiguities and circular arguments in terms only of the mechanical concepts of space, time, and force or inertial mass.

The order of introduction of concepts and principles is system (types and amounts of constituents, forces between constituents, and external forces or parameters); properties; states; first law (without energy, work, and heat); energy (without work and heat); energy balance; classification of states in terms of time evolutions; stable equilibrium states; second law (without temperature, heat, and entropy); generalized available energy; entropy of any state (stable equilibrium or not) in terms of energy and generalized available energy, and not in terms of temperature and heat; entropy balance; fundamental relation for stable equilibrium states only; temperature, total potentials (chemical and/or electrochemical), and pressure in terms of energy, entropy, amounts of constituents and parameters for stable equilibrium states only; third law; work in terms of energy; and heat in terms of energy, entropy, and temperature.

All concepts and principles are valid for all systems (both macroscopic, and microscopic), all states (both thermodynamic or stable equilibrium states, and states that are not stable equilibrium), and involve no statistical probabilities (neither statistical classical mechanics nor statistical quantum mechanics).

**A.1.2 Definition**

We define general thermodynamics or simply thermodynamics as the study of motions of physical constituents (particles and radiations) resulting from externally applied forces, and from internal forces (the actions and reactions between constituents). This definition is identical to that given by Timoshenko and Young about mechanical dynamics [26]. However, because of the second law, we will see that the definition encompasses a much broader spectrum of phenomena than mechanical dynamics.

**A.1.3 Kinematics: conditions at an instant in time**

In kinematics we give verbal definitions of the terms system, property, and state so that each definition is valid without change in any paradigm of physics, and involves no statistics attributable to lack of information about any aspect of a problem in physics, and/or consideration of numerical and computational difficulties. The definitions include innovations. To the best of our knowledge, they violate no theoretical principle, and contradict no experimental results.

A *system* is defined as a collection of *constituents*, subject to *internal forces*, that is, forces between constituents, and *external forces*, that is, forces that depend only on coordinates of the constituents and not on coordinates of constituents of the source of the force. Everything that is not included in the system is the *environment*. For these definitions to be meaningful, the system must be both separable from and uncorrelated with any entity in its environment.

For a system with r constituents, subject to external forces described by s parameters, we denote the amounts by the vector $\boldsymbol{n} = \{n_1, n_2, \ldots, n_r\}$, and the parameters by the vector $\boldsymbol{\beta} = \{\beta_1, \beta_2, \ldots, \beta_s\}$. One parameter may be volume, $\beta_1 = V$, another may be an externally determined electric field, $\beta_2$ equal to the electric field. At any instant in time, the amount $n_i$ of constituent i and the parameter $\beta_j$ of external force j have specific values. We denote all such values by $\boldsymbol{n}$ and $\boldsymbol{\beta}$ with or without additional subscripts.

By themselves, the values of the amounts of constituents and of the parameters at an instant in time do not suffice to characterize completely the condition of the system at that time. We also need the values of a complete set of linearly independent properties at the same instant in time. A *property* is defined as an attribute that can be evaluated at any given instant in time (not as an average over time) by means of a set of measurements and operations that are performed on the system and result in a numerical value – *the value of the property*. This value is independent of the measuring devices, other systems in the environment, and other instants in time.

For a given system, the instantaneous values of the amounts of all the constituents, the values of all the parameters, and the values of a complete set of linearly independent properties encompass all that can be said about the system at a given instant in time, and about the results of any measurements that may be performed on the system at that same instant in time. We call this complete characterization of the system at any instant in time the *state* of the system. This definition of state, without change, applies to all paradigms of physics.

**A.1.4 Dynamics: changes of state in time**

The state of a system may change in time either spontaneously due to the internal forces or as a result of interactions with other systems, or both. The relation that describes the evolution of the state of an isolated system – *spontaneous changes of state* – as a function of time is the *equation of motion*. Certain time evolutions obey *Newton's equation* which relates the force $F$ on the constituents of the system to their inertial mass *m* and acceleration *a* so that $F = ma$. Other evolutions obey the *Schrödinger equation*, that is, the quantum-mechanical equivalent of Newton's equation. Other experimentally observed time evolutions, however, do not obey either of these equations. So the equations of motion that are widely known are correct but incomplete. The discovery of the complete equation of motion that describes all quantum mechanical evolutions in time is a subject of research and controversy at the frontier of science – one of the most intriguing and challenging problems in quantum physics [18-20, 27-30]. To the best of our knowledge, the only complete equations, that satisfy all known requirements are given in Refs. [19, 20].

Two of the most general and well-established features of the complete equation of motion are captured by the consequences of the first and second laws of thermodynamics discussed in the next sections.

## A.2 Energy

### A.2.1 The first law, energy and energy balance

Energy is a concept that underlies our understanding of all physical phenomena, yet its meaning is subtle and difficult to grasp. It emerges from a fundamental principle known as the first law of thermodynamics.

*The first law asserts that any two states of a system may always be the initial and final states of a weight process. Such a process involves no net effects external to the system except the change in elevation between $z_1$ and $z_2$ of a weight, that is, solely a mechanical effect. Moreover, for a given weight, the value of the expression $M g (z_1 - z_2)$ is fixed only by the end states of the system, where M is the mass of the weight, and g the gravitational acceleration.*

One theorem of this law is that every system *A* in any state $A_1$ has a property called *energy*, with a value denoted by the symbol $E_1$. The energy $E_1$ can be evaluated by a weight process that connects $A_1$ and a reference state $A_0$ to which is assigned an arbitrary reference value $E_0$ so that

$$E_1 = E_0 - M g (z_1 - z_0) \tag{A-1}$$

Energy is shown to be an additive property, that is, the energy of a composite of two or more identifiable subsystems is the sum of the energies of the subsystems. Moreover, it is also shown that energy has the same value at the final time as at the initial time if the system experiences a zero-net-effect weight process, and that energy remains invariant in time if the process is spontaneous. In either of the last two processes, $z_2 = z_1$ and $E(t_2) = E(t_1)$ for time $t_2$ greater than $t_1$, that is, energy is *conserved*. Energy conservation is a time-dependent result. In Ref. [22], this result is obtained without use of the complete equation of motion.

Energy is transferred between systems as a result of interactions. Denoting by $E^{A \leftarrow}$ the amount of energy transferred from the environment to system *A* in a process that changes the state of *A* from $A_1$ to $A_2$, we can derive the *energy balance*. This derivation is based on the additivity of energy and energy conservation, and reads

$$(E_2 - E_1)_{\text{system } A} = E^{A \leftarrow} \tag{A-2}$$

In words, the energy change of a system must be accounted for by the energy transferred across the boundary of the system.

## A.3 Available Energy

### A.3.1 Types of states

Because in quantum theory the number of independent properties of a system is infinite even for a system consisting of a single particle with a single translational degree of freedom – a single variable that fixes the configuration of the system in space – and because most properties can vary over a range of values, the number of possible states of a system is infinite. The discussion of these states is facilitated if they are classified into different categories according to evolutions in time. This classification brings forth many important aspects of physics, and provides a readily understandable motivation for the introduction of the second law of thermodynamics.

The classification consists of unsteady states (they change in time as a result of interactions),

steady states (they do not change in time despite interactions), nonequilibrium states, and equilibrium states. An *equilibrium state* is one that does not change in time while the system is isolated—a state that does not change spontaneously. An *unstable equilibrium state* is an equilibrium state that may be caused to proceed spontaneously to a sequence of entirely different states by means of a minute and short-lived interaction that has either an infinitesimal effect or a zero net effect on the state of the environment. A *stable equilibrium state* is an equilibrium state that can be altered to a different state only by interactions that leave net effects in the environment of the system. These definitions are identical to the corresponding definitions in mechanics but include a much broader spectrum of states than those encountered in mechanics. The broader spectrum is due to the second law discussed later.

Starting either from a nonequilibrium state or from an equilibrium state that is not stable, experience shows that energy can be transferred out of the system, and affect a mechanical effect without leaving any other net changes in the state of the environment. In contrast, starting from a stable equilibrium state, experience shows that energy cannot be transferred out of the system while achieving only the mechanical effect just cited. This impossibility is one of the most striking consequences of the first and second laws of thermodynamics.

**A.3.2  The second law and generalized adiabatic availability**

The existence of stable equilibrium states is not self-evident. It was recognized by Hatsopoulos and Keenan [31] as the fundamental theoretical underpinning of all correct statements of the second law in practically every textbook on either thermodynamics or physics. In essence, these statements are theorems arising from the concept of stability. Gyftopoulos and Beretta [21] concur with this recognition, and state the *second law* as follows (simplified version): *Among all the states of a system with a given value of energy, and given values of the amounts of constituents and the parameters, there exists one and only one stable equilibrium state*.

The existence of stable equilibrium states for the conditions specified and therefore the second law cannot be derived from the laws of mechanics. Within mechanics, the stability analysis yields that among all the allowed states of a system with fixed values of amounts of constituents and parameters, the only globally stable equilibrium state is that of lowest energy. In contrast the second law avers the existence of a much larger class of globally stable equilibrium states in addition to the states contemplated by mechanics.

The existence of stable equilibrium states for various conditions of matter has many theoretical and practical consequences. One consequence (theorem) is that, starting from a stable equilibrium state of any system, no energy is available to affect a mechanical effect while the values of the amounts of constituents and parameters of the system experience no net changes. This consequence is often referred to as the impossibility of the perpetual motion machine of the second kind (PMM2). In many expositions of thermodynamics, it is taken as the statement of the second law. In our exposition, it is only one theorem of both the first and the second laws.

Another consequence is that not all states of a system can be changed to a state of lowest energy by means of a mechanical effect. This is a generalization of the impossibility of a PMM2. In essence, it is shown that a novel important property exists which is called *generalized adiabatic availability*. The generalized adiabatic availability of a system in a given state represents the optimum amount of energy that can be exchanged between the system and a weight in a weight process. Like energy, this property is well defined for all systems and all states, but unlike energy it is not additive.

**A.3.3  Generalized available energy**

In striving to define an additive property that captures the important features of generalized adiabatic availability, Gyftopoulos and Beretta introduce a special reference system, called a

r*eservoir* (see definition in Sec. 2.1), and discuss the possible weight processes that the composite of a system and a reservoir may experience. Thus they disclose a third consequence of the first and second laws, that is, a limit on the optimum amount of energy that can be exchanged between a weight and a composite of a system and a reservoir $R$—the optimum mechanical effect. They call the optimum value *generalized available energy*, denote it by $\Omega^R$, and show that it is additive. It is a generalization of the concept of motive power of fire introduced by Carnot. It is a generalization because he assumed that both systems of the composite acted as reservoirs with fixed values of their respective amounts of constituents and parameters, whereas Gyftopoulos and Beretta do not use this assumption.

For an *adiabatic process* of system $A$ only, it is shown that the energy change $E_1 - E_2$ of $A$, and the generalized available energy change $\Omega_1^R - \Omega_2^R$ of the composite of $A$ and reservoir $R$ satisfy the following relations. If the adiabatic process of $A$ is reversible,

$$E_1 - E_2 = \Omega_1^R - \Omega_2^R \tag{A-3}$$

If the adiabatic process of $A$ is irreversible,

$$E_1 - E_2 < \Omega_1^R - \Omega_2^R \tag{A-4}$$

A process is *reversible* if both the system and its environment can be restored to their respective initial states. A process is *irreversible* if the restoration just cited is impossible. In each of these restorations, the sequence of restoring states need not, and in practically all processes is not a retrace of the initial sequence. Excellent examples of these facts are the reversible or irreversible processes involved in a Carnot cycle. In addition, such considerations bring forth the question of whether there is an arrow of time, but the discussion of this question is beyond the scope of this presentation.

It is noteworthy that energy and generalized available energy are defined for any state of any system, regardless of whether the state is steady, unsteady, equilibrium, nonequilibrium, or stable equilibrium, and regardless of whether the system has many degrees of freedom or one degree of freedom, or whether the size of the system is large or small.

It is also noteworthy that the laws of thermodynamics do not require that either Eq. A-3 or Eq. A-4 be always true. They simply stipulate one of the consequences of either reversibility or irreversibility.

**A.4 ENTROPY**

**A.4.1 Definition**

A system $A$ in any state $A_1$ has many properties. Two of these properties are energy $E_1$ and generalized available energy $\Omega_1^R$ with respect to a given auxiliary reservoir $R$. These two properties determine a third property called *entropy*, denoted by the symbol $S$. It is a property in the same sense that energy is a property, or momentum is a property. For a state $A_1$, an auxiliary reservoir $R$, a reference state $A_0$, with energy $E_0$, generalized available energy $\Omega_0^R$, and a reference entropy $S_0$, the entropy $S_1$ of $A_1$ is defined by the relation

$$S_1 = S_0 + \frac{1}{c_R}[(E_1 - E_0) - (\Omega_1^R - \Omega_0^R)] \tag{A-5}$$

where $c_R$ is a well-defined positive constant that depends on the auxiliary reservoir $R$ only. Entropy $S$ is shown to be independent of the reservoir, that is, $S$ is a property of system $A$ only, and $R$ is auxiliary. The reservoir is used only because it facilitates the definition of $S$. It is also shown that $S$ can be assigned absolute values that are non-negative, and that vanish for all the states encountered in classical or conventional quantum mechanics. Because both energy and generalized available energy are additive, entropy is also additive.

**A.4.2 Reversible and irreversible processes**

Because energy and generalized available energy satisfy relations (A-3) and (A-4), the entropy defined by Eq. (A-5) remains invariant in any reversible adiabatic process of $A$, and increases in any irreversible adiabatic process of $A$. These conclusions are valid also for spontaneous processes, and for zero-net-effect interactions. The latter features are known as *the principle of nondecrease of entropy*. Both a spontaneous process and a zero-net-effect interaction are special cases of an adiabatic process of system *A*.

The entropy created as a system proceeds from one state to another during an irreversible process is called *entropy generated by irreversibility*. It is positive. The entropy nondecrease is a time-dependent result. In the novel exposition of thermodynamics [22], this result is obtained without use of the complete equation of motion.

Like energy, entropy can be transferred between systems by means of interactions. Denoting by $S^{A\leftarrow}$ the amount of entropy transferred from systems in the environment to system *A* as a result of all interactions involved in a process in which the state of *A* changes from state $A_1$ to state $A_2$, we derive a very important analytical tool, the *entropy balance*, that is,

$$(S_2 - S_1)_{\text{system } A} = S^{A\leftarrow} + S_{\text{irr}}, \qquad (A-6)$$

where $S_{\text{irr}}$ is non-negative. A positive $S_{\text{irr}}$ represents the entropy generated spontaneously within system *A* in the time interval from $t_1$ to $t_2$ required to affect the change from state $A_1$ to state $A_2$. Spontaneous entropy generation within a system occurs if the system is in a nonequilibrium state because of the natural tendency of the system to reach an equilibrium or a stable equilibrium state.

The dimensions of *S* depend on the dimensions of both energy and $c_R$. It turns out that the dimensions of $c_R$ are independent of mechanical dimensions, and are the same as those of temperature. Temperature is defined later.

### A.4.3 Properties of stable equilibrium states

It is shown that among the many states of a system that have given values of energy *E*, amounts of constituents *n*, and parameters *β*, the entropy of the unique stable equilibrium state that corresponds to these values is larger than that of any other state with the same values *E*, *n*, and *β*, and can be expressed as a function

$$S = S(E, n, \beta) \qquad (A-7)$$

Equation (A-7) is called the *fundamental relation*.

The fundamental relation is shown to be analytic in each of its variables *E*, *n*, and *β*, and concave with respect to energy, that is,

$$(\partial^2 S / \partial^2 E)_{n,\beta} \leq 0 \qquad (A-8)$$

Moreover, the fundamental relation is used to define other properties of stable equilibrium states, such as *temperature T*

$$1/T = (\partial S / \partial E)_{n,\beta}, \qquad (A-9)$$

*total potentials* $\mu_i$

$$\mu_i = -T(\partial S / \partial n_i)_{E, n, \beta} \quad \text{for} \quad i = 1, 2, \ldots, r, \quad (A-10)$$

and *pressure p*

$$p = T(\partial S / \partial V)_{E, n, \beta} \qquad (A-11)$$

The temperature, total potentials, and pressure of a stable equilibrium state appear in the necessary and sufficient conditions for systems to be in mutual stable equilibrium, such as the temperature equality, the total potential equality, and the pressure equality. Moreover, these equalities are the bases for the measurements of *T*, $\mu_i$'s, and *p*.

### A.4.4 The third law

For a system without an upper bound on energy, the third law of thermodynamics asserts that: *for each set of values of the amounts of constituents and the parameters, there exists one stable equilibrium state with temperature T = 0 or, equivalently,* $1/T = \infty$. For a system with an upper bound on energy, such as a spin, the third law asserts that: *there exists two stable equilibrium states with T = 0, one with $1/T = \infty$, and the other with $1/T = -\infty$*.

It is noteworthy that in the unified quantum theory of mechanics and thermodynamics the three laws of thermodynamics are theorems of the theory in the same sense that momentum conservation and kinetic energy conservation in elastic collisions in classical mechanics are theorems of *F = ma*.

### 4.5 CHARACTERISTICS OF ENTROPY

From the discussions in the preceding section and our knowledge of classical thermodynamics, we conclude that any expression that purports to represent the entropy *S* of thermodynamics must have at least the following eight characteristics or, equivalently, conform to the following eight criteria.
(1) *S* must be well defined for every system (large or small), and every state (stable equilibrium or not stable equilibrium).
(2) *S* must be invariant in all reversible adiabatic processes, and increase in any irreversible adiabatic process.
(3) *S* must be additive for all systems and all states.
(4) *S* must be non-negative, and vanish for all the states encountered in mechanics.
(5) For given values of energy, amounts of constituents, and parameters, one and only one state must correspond to the largest value of *S*.
(6) For given values of the amounts of constituents and parameters, the graph of entropy versus energy of stable equilibrium states must be concave and smooth.
(7) For a composite *C* of two subsystems *A* and *B*, the expression must be such that the maximization of *S* of *C* [criterion no. (5)] yields identical thermodynamic potentials (for example, temperature, total potentials, and pressure) for all three systems *A*, *B*, and *C*.
(8) For stable equilibrium states, *S* must reduce to relations that have been established experimentally and that express the entropy in terms of values of energy, amounts of constituents, and parameters, such as the relations for ideal gases.

It is noteworthy that, except for criteria (1) and (4), we can establish the remaining six criteria by reviewing the definition of entropy of classical thermodynamics.

### 4.6 COMMENT

The concept of entropy introduced here differs from and is more general than what is presented in practically all textbooks on thermodynamics and / or physics. It does not involve the concepts of temperature and heat; it is not restricted to large systems; it applies to macroscopic as well as microscopic systems, including a system with one spin, or a system with one particle with only one (translational) degree of freedom; it is not restricted to stable (thermodynamic) equilibrium states; it is defined for both stable equilibrium (thermodynamic equilibrium) and not stable equilibrium states because energy and generalized available energy are defined for all states; and most certainly, it is not statistical – it is an intrinsic property of matter. These assertions are valid because here the postulates or laws of thermodynamics from which the concept of entropy originates do not involve the concepts of temperature and heat, are not restricted either to large systems or to stable equilibrium states or both, and are not statistical. To emphasize the difference and generality of the concept, we recall contrary statements by Meixner [32]: "A careful study of the thermodynamics of electrical networks has given considerable insight into these problems and also produced a very interesting result: the nonexistence of a unique entropy value in a state which is obtained during an irreversible process,…, I would say I have done away with entropy"; by Callen [33]: "It must be stressed that we postulate the existence of the entropy only for equilibrium states and that our postulate makes no reference whatsoever to nonequilibrium states", and by Lieb and Yngvason [34] "Once again, it is a good idea to try to understand first the meaning of entropy for equilibrium states – the quantity that our textbooks talk about when they draw Carnot cycles. In this article we restrict our attention to just those states". It is noteworthy, that even from their totally unjustified limited perspective, Lieb and Yngvason [35] introduce 16 axioms to explain their second law of thermodynamics(!).